\documentclass[aps,prl,twocolumn,showpacs,superscriptaddress,floatfix]{revtex4-1}
\usepackage{amsmath,amssymb}
\usepackage{graphicx}
\usepackage{epsfig}
\usepackage{color}
\usepackage{rotating}
\usepackage{bm}
\usepackage{setspace}
\usepackage{hyperref}

\begin{document}
\title{High-Temperature Superconductivity in Single-Unit-Cell FeSe Films \\on Anatase TiO$_2$(001)}
\author{Hao Ding}
\author{Yan-Feng Lv}
\author{Kun Zhao}
\author{Wen-Lin Wang}
\affiliation{State Key Laboratory of Low-Dimensional Quantum Physics, Department of Physics, Tsinghua University, Beijing 100084, China}
\author{Lili Wang}
\author{Can-Li Song}
\author{Xi Chen}
\author{Xu-Cun Ma}
\author{Qi-Kun Xue}
\email[]{qkxue@mail.tsinghua.edu.cn}
\affiliation{State Key Laboratory of Low-Dimensional Quantum Physics, Department of Physics, Tsinghua University, Beijing 100084, China}
\affiliation{Collaborative Innovation Center of Quantum Matter, Beijing 100084, China}

\begin{abstract}
We report on the observation of high-temperature ($T_\textrm{c}$) superconductivity and magnetic vortices in single-unit-cell FeSe films on anatase TiO$_2$(001) substrate by using scanning tunneling microscopy. A systematic study and engineering of interfacial properties has clarified the essential roles of substrate in realizing the high-$T_\textrm{c}$ superconductivity, probably via interface-induced electron-phonon coupling enhancement and charge transfer. By visualizing and tuning the oxygen vacancies at the interface, we find their very limited effect on the superconductivity, which excludes interfacial oxygen vacancies as the primary source for charge transfer between the substrate and FeSe films. Our findings have placed severe constraints on any microscopic model for the high-$T_\textrm{c}$ superconductivity in FeSe-related heterostructures.
\end{abstract}

\maketitle
\begin{spacing}{1.04}
The recent discovery of superconductivity with an exceptionally high critical temperature ($T_\textrm{c}$) over 65 K in single-unit-cell (SUC) FeSe films on SrTiO$_3$ has received extensive attention \cite{qing2012interface, liu2012electronic, he2013phase, tan2013interface, xiang2012high, bang2013atomic, lee2014interfacial, miyata2015high, rademaker2015enhanced, li2015quantum}. Distinct from iron pnictide and bulk FeSe compounds \cite{Nakayama2014reconstruction, song2011direct}, the superconducting SUC FeSe films prepared on SrTiO$_3$ substrate not only possess a rather simple Fermi surface topology-only having electron pockets around the zone corner (\textit{M} point) of Brillouin zone Fermi surface ($E_F$) \cite{liu2012electronic, tan2013interface, lee2014interfacial}, but also reach a record of $T_\textrm{c}$ values among all iron-based superconductors (Fe-SCs) \cite{qing2012interface, liu2012electronic, he2013phase, tan2013interface}. In order to understand the enhancement in $T_\textrm{c}$, several different scenarios invoking interface effects, such as interfacial electron-phonon coupling \cite{qing2012interface, lee2014interfacial, rademaker2015enhanced, li2015quantum}, charge transfer prompted by interfacial oxygen vacancies \cite{liu2012electronic, he2013phase, tan2013interface, bang2013atomic}, tensile strain effect induced by the lattice mismatch between FeSe and SrTiO$_3$ \cite{tan2013interface, peng2014tuning}, screening effect by SrTiO$_3$ ferroelectric phonons \cite{xiang2012high}, have been proposed. However, a consensus on which factors play the primary roles in enhancing the superconductivity of FeSe has not yet be achieved. The situation is further complicated by the observation of high-$T_\textrm{c}$ superconductivity in anisotropic SrTiO$_3$(110) substrates \cite{zhou2015observation, zhang2015observation} and heavily electron-doped FeSe-derived superconductors involving little interfacial effect \cite{burrard2013enhancement, lu2014coexistence, lei2015evolution, Song2016observation, zhang2015superconducting}.

Attempt to separate the effects of substrate by preparing FeSe films on graphitized SiC(0001) leads to the discovery of two disconnected superconducting domes upon alkali-metal potassium doping \cite{Song2016observation}: a low-$T_\textrm{c}$ phase in undoped parent FeSe and a high-$T_\textrm{c}$ phase in heavily electron-doped regime. This points out a different pairing mechanism of high-$T_\textrm{c}$ phase from that in other Fe-SCs. Most importantly, the observed $T_\textrm{c}$ of $\sim$ 48 K and $\Delta$ of $\sim$ 14 meV in these systems \cite{burrard2013enhancement, lu2014coexistence, Song2016observation, zhang2015superconducting} imply that, to boost the higher $T_\textrm{c}$ in FeSe \cite{qing2012interface, liu2012electronic, he2013phase, tan2013interface}, the SrTiO$_3$ must impose additional effect(s) other than the electron doping. In order to clarify the above-mentioned controversies, it is highly desirable to find an alternative substrate that can host high-$T_\textrm{c}$ superconductivity and meanwhile allow a straightforward comparison with SrTiO$_3$.

Realizing that the SrTiO$_3$ substrates commonly have a termination of TiO$_2$ under ultrahigh vacuum (UHV) preparation conditions, we grow anatase TiO$_2$ directly on SrTiO$_3$ as substrate for FeSe growth by molecular beam epitaxy (MBE). The epitaxial anatase TiO$_2$ with various distinctive physical properties from SrTiO$_3$ allows for a direct scrutiny of interfacial effects on the emerging high-$T_\textrm{c}$ superconductivity.

Our experiments were conducted in a Unisoku UHV scanning tunneling microscopy (STM) system, equipped with a MBE chamber for film preparation. The base pressure is better than 1.0 $\times$ 10$^{-10}$ Torr. The 0.05 wt Nb-doped SrTiO$_3$(001) substrates were degassed at 600$^\textrm{o}$C for 3 hours, and then annealed at 1250$^\textrm{o}$C for 20 minutes to get clean and flat surface. Anatase TiO$_2$ islands were grown on SrTiO$_3$(001) at 750$^\textrm{o}$C by evaporating high-purity Ti (99.99$\%$). At this temperature, oxygen atoms decompose from SrTiO$_3$ and react with the deposited Ti to form anatase TiO$_2$. The growth rate was approximately 0.11 monolayers per minute, which was calibrated by STM. FeSe films were then grown on anatase TiO$_2$(001) surface at 500$^\textrm{o}$C by co-evaporating high-purity Fe (99.995$\%$) and Se (99.999$\%$) from standard Knudsen cells under Se-rich condition. The growth rate was $\sim$ 0.01 layers per minute. With the growth recipe, the as-grown SUC FeSe films on anatase TiO$_2$(001) are superconducting without need of post-growth annealing. All STM images were acquired at 4.2 K with a polycrystalline PtIr tip. The \textit{dI/dV} spectra were acquired using lock-in technique with a bias modulation of 10 mV and 0.5 mV for wide-energy-scale (-0.5 V $\sim$ 0.5 V) and small-energy-scale (-50 mV $\sim$ 50 mV) spectra at 913 Hz, respectively.
\end{spacing}

\begin{figure}[h]
\includegraphics[width=\columnwidth]{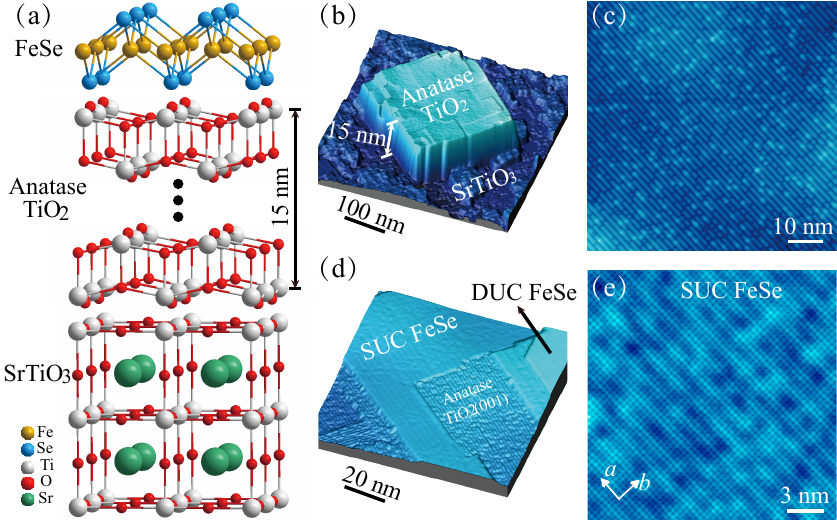}
\caption{(color online) (a) Schematic of FeSe/TiO$_2$(001) heterostructure. The surface reconstructions of SrTiO$_3$(001) and anatase TiO$_2$(001) are not shown for clarity. For the typical TiO$_2$(001) thick films studied, its in-plane lattice constant has been well relaxed to 0.380 nm from 0.3905 nm. (b) STM topography (500 nm $\times$ 500 nm, $V_\textrm{s}$ = 3.0 V, $I_\textrm{t}$ = 0.03 nA) showing an anatase TiO$_2$(001) island with a thickness of 15 nm supported by a SrTiO$_3$(001) substrate. (c) Zoom-in STM topography (70 nm $\times$ 70 nm, $V_\textrm{s}$ = 1.5 V, $I_\textrm{t}$ = 0.03 nA) acquired on a TiO$_2$ island, showing clear 4 $\times$ 1 reconstruction plus oxygen vacancies (bright spots). (d) STM topography (100 nm $\times$ 100 nm, $V_\textrm{s}$ = 3.0 V, $I_\textrm{t}$ = 0.03 nA) showing coexisting SUC and DUC FeSe films on anatase TiO$_2$(001). (e) Atomically resolved STM topography (20 nm $\times$ 20 nm, $V_\textrm{s}$ = 50 mV, $I_\textrm{t}$ = 0.1 nA) of SUC FeSe film. \textit{a} and \textit{b} correspond to either of Se-Se nearest-neighbor directions throughout.
}
\end{figure}

\begin{figure*}[t]
\includegraphics[width=2\columnwidth]{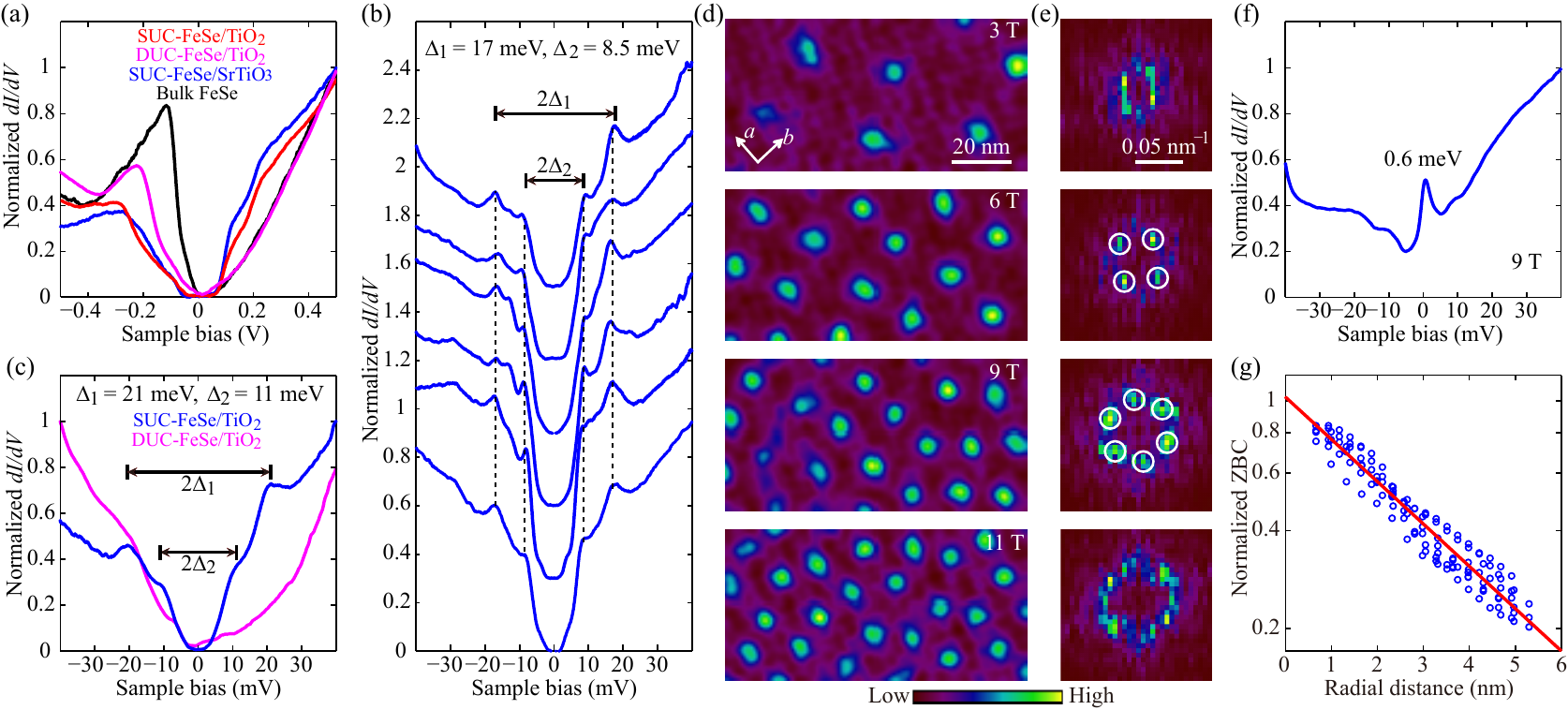}
\caption{(color online) (a) Comparison of large-energy-scale \textit{dI/dV} spectra (set point: $V_\textrm{s}$ = 0.5 V, $I_\textrm{t}$ = 0.1 nA) of FeSe films on anatase TiO$_2$(001) (red and magenta curves) and SrTiO$_3$(001) (blue curve), as well as undoped bulk FeSe (black curve). (b, c) Low energy \textit{dI/dV} spectra (set point: $V_\textrm{s}$ = 50 mV, $I_\textrm{t}$ = 0.4 nA) taken on various SUC FeSe/TiO$_2$(001) films (blue curves) revealing the typical and largest superconducting gap, respectively. Vertical dashes are guide to eye. The spectrum color-coded by magenta in c indicates no superconductivity on DUC FeSe films. (d) ZBC map (100 nm $\times$ 50 nm, set point: $V_\textrm{s}$ = 50 mV, $I_\textrm{t}$ = 0.1 nA) showing the vortices under various magnetic fields in SUC FeSe/TiO$_2$(001). Yellow regions with enhanced ZBC, due to the suppressed superconducting gap, indicate the individual isolated vortices. (e) FFT power spectra of the ZBC maps in d. (f) Differential conductance \textit{dI/dV} spectrum (set point: $V_\textrm{s}$ = 50 mV, $I_\textrm{t}$ = 0.4 nA) measured at a vortex center at 9 T, showing the suppressed superconductivity and Andreev bound states at 0.6 meV. (g) Radial dependence of normalized ZBC around a single vortex core (blue circles). The fit to an exponential decay (red line), namely ZBC(\textit{r}) = $g_{0}$($x$ = $\infty$) + $A$exp(-\textit{r}/$\xi_{\textrm{GL}}$) with $g_{0}$($x$ = $\infty$) as the constant background, leads to an angularly averaged GL coherence length $\xi_{\textrm{GL}}$ = 2.85 $\pm$ 0.14 nm. The normalized \textit{dI/dV} spectrum was obtained by dividing the raw one by its conductance value at the tunneling set point.}
\end{figure*}

As sketched in Fig.\ 1(a), anatase TiO$_2$ is characterized with distinct O-Ti-O triple-layered planes in sharp contrast to the single-layered TiO$_2$ planes in SrTiO$_3$. This has consequently led to a great deal of diversities between their physical properties such as lattice constants, phonon modes and dielectric constants. The in-plane lattice constant of anatase TiO$_2$(001) is 0.3782 nm, much closer to that (0.3765 nm) of FeSe than that (0.3905 nm) of SrTiO$_3$(001). Distinct from a ferroelectric soft phonon mode with a frequency of 88 cm$^{-1}$ in SrTiO$_3$, which has been proposed to be crucial for the $T_\textrm{c}$ enhancement in SUC FeSe/SrTiO$_3$ \cite{xiang2012high}, the corresponding phonons in anatase TiO$_2$ have a quite high frequency (367 cm$^{-1}$) and a very limited contribution to static dielectric constant (22.7 along the \textit{c}-axis, significantly smaller than 300 of SrTiO$_3$) \cite{Mikami2002lattice}. As thus, the anatase TiO$_2$(001) serves as an advisory system to distinguish whether or not the above-mentioned parameters bear a primary responsibility for the enhanced high-$T_\textrm{c}$ superconductivity in FeSe-related heterostructures. Furthermore, in contrast with SrTiO$_3$ where the oxygen vacancies have never been directly identified \cite{qing2012interface, zhou2015observation, zhang2015observation}, oxygen vacancies on anatase TiO$_2$ can be easily tuned in their density by annealing \cite{wang2013role} and visualized by STM. This renders anatase TiO$_2$ a unique substrate for a systematic study of the role of oxygen vacancies in the superconductivity of SUC FeSe.

The as-grown samples exhibit isolated TiO$_2$ islands with a typical thickness of 15 nm and a lateral size as large as 300 nm $\times$ 300 nm on SrTiO$_3$(001) [Fig.\ 1(b)]. Figure 1(c) depicts a zoom-in image of TiO$_2$(001) island, which clearly reveals the well-known 4 $\times$ 1 reconstruction decorated by some oxygen vacancies (bright spots) \cite{wang2013role}. After a precise calibration of our STM piezotube scanner, the in-plane lattice constant is measured to be 0.380 $\pm$ 0.005 nm (close to the value of 0.3782 nm in bulk TiO$_2$), indicative of a nearly fully relaxed lattice. On the substrate with the unstrained anatase TiO$_2$(001) islands, we then grow FeSe films using the previously established method \cite{qing2012interface, song2011direct}, which leads to formation of atomically flat SUC and double-unit-cell (DUC) films with very few defects [Figs.\ 1(d) and 1(e)]. Our subsequent STM measurements reveal the same in-plane lattice constant of both SUC and DUC FeSe as TiO$_2$, 0.380 $\pm$ 0.005 nm, slightly larger than the value (0.3765 nm) of bulk FeSe within the experimental uncertainty. The result suggests nearly strain-free FeSe films formed on anatase TiO$_2$(001), resembling with those on graphene/SiC(0001) \cite{song2011direct, Song2016observation}.

Tunneling conductance \textit{dI/dV} spectrum on SUC (red curve) and DUC (magenta curve) FeSe films on anatase TiO$_2$(001) are illustrated in Fig.\ 2(a), which behave very differently. Although the overall feature on DUC resembles more closely with undoped parent FeSe (black curve), the \textit{dI/dV} spectrum on SUC differs markedly from both, but bears great similarities with those acquired in SUC FeSe films on SrTiO$_3$ (blue curve) and in K-doped FeSe films \cite{Song2016observation}. The result reveals a substantial electron transfer from anatase TiO$_2$(001) to SUC FeSe film. Subsequent \textit{dI/dV} measurements in a smaller energy scale [Figs.\ 2(b) and 2(c)] reveal first evidence of superconductivity as SUC FeSe/SrTiO$_3$: spatially rather universal and U-shaped gaps with vanishing density of states near the Fermi energy ($E_F$). The double-gap (denoted as $\Delta_1$ and $\Delta_2$, respectively) structure most probably suggests a multi-band superconductor \cite{zhang2015superconducting, fan2015plain}. The maximal gap $\Delta_1$ varies from 17 meV [Fig.\ 2(b)] to 21 meV [Fig.\ 2(c)], even slightly larger than those measured in SUC FeSe/SrTiO$_3$ films \cite{qing2012interface, liu2012electronic, he2013phase, tan2013interface, lee2014interfacial, fan2015plain}. This might mean a higher $T_\textrm{c}$ in SUC FeSe/TiO$_2$ films, which merits a further study of the temperature-dependent superconducting gap in the future. Note that the DUC FeSe films on TiO$_2$(001) exhibit no superconductivity signature at all [Fig.\ 2(c)], resembling with the DUC FeSe films on SrTiO$_3$ \cite{qing2012interface}. This may be caused by a small amount of but insufficient electron transfer from TiO$_2$ to DUC FeSe films, which pushes them to the nonsuperconducting region between the recently discovered two superconducting domes of FeSe \cite{Song2016observation}.

To confirm that the gap opening around $E_F$ observed in Figs.\ 2(b) and 2(c) is related to superconductivity, we have conducted varying magnetic field experiment. Application of magnetic field perpendicular to a superconducting SUC FeSe/TiO$_2$(001) films can suppress its local superconductivity, leading to the formation of vortices and Abrikosov lattice. This situation is demonstrated by the zero-bias conductance maps in Fig.\ 2(d). The magnetic vortices, seen as zero-bias conductance enhancement, increase linearly in number with the field \textit{B}, as anticipated. Remarkably a nearly square vortex lattice, more clearly seen in fast Fourier-transferred (FFT) images in Fig.\ 2(e), is observable at \textit{B} = 6 T. This supports a fourfold anisotropy in the superconducting order parameter \cite{Wilde1997scanning, Nakai2002reentrant, Brown2004triangular}, bearing a strong likeness to the moderately anisotropic gap in SUC FeSe/SrTiO$_3$ films measured by a recent angle-resolved photoemission spectroscopy experiment \cite{zhang2015superconducting}. Based on this model, the gap minima should be located along the nearest-neighbor directions of the square vortex lattice, which is orientated along the Se-Se bond directions [Fig.\ 2(d)]. This observation puts strong constraints on the electron pairing symmetry in SUC FeSe/TiO$_2$(001) films. At 9 T, the vortices change into a distorted triangle lattice. Such evolution of vortex structure has previously been found in YNi$_2$B$_2$C \cite{Imaging2000Sakata}, high-$T_\textrm{c}$ cuprate La$_{1.83}$Sr$_{0.17}$CuO$_4$ \cite{Gilardi2002direct} and CeCoIn$_5$ \cite{zhou2013visualizing}, which might be caused by other competing effects, such as Fermi surface anisotropy \cite{reentrant2012Nakai}.

In the vicinity of vortex core, our \textit{dI/dV} measurements reveal the presence of an Andreev bound state at $E_0$ = 0.6 meV linking with lowest bound state \cite{Hayashi1998low} and the complete disappearance of double-superconducting-gap structure [Fig.\ 2(f)]. The results demonstrate unambiguously the gaps as superconducting gaps as well as the occurrence of high-$T_\textrm{c}$ superconductivity in SUC FeSe/TiO$_2$(001) films. Analysis of the spatial evolution of radial zero-bias-conductance (ZBC) values around the vicinity of vortices [Fig.\ 2(g)] has led to a Ginzburg-Landau (GL) coherence length $\xi_{\textrm{GL}}$ = 2.85 $\pm$ 0.14 nm, consistent with that (2.45 $\sim$ 3.18 nm) of SUC FeSe/SrTiO$_3$ films \cite{fan2015plain}. The observed particle-hole asymmetry at the vortex core with the lowest bound state on the empty state [Fig.\ 2(f)] indicates electron-type charge carriers in superconducting SUC FeSe/TiO$_2$(001) \cite{Hayashi1998low}, in good agreement with the \textit{dI/dV} measurements above [Fig.\ 2(a)].

Our demonstration of superconductivity with a very large $\Delta$ in the SUC FeSe films on TiO$_2$ constitutes a critical finding related to the interfacial high-$T_\textrm{c}$ superconductivity. The finding, that $\Delta$ appears larger than those (8 $\sim$ 14 meV) in other heavily electron-doped FeSe-derived superconductors with no interfacial effect \cite{Song2016observation, zhang2015superconducting}, has recalled the crucial role of interfacial effects in the occurrence of high-$T_\textrm{c}$ superconductivity. A direct comparison with SUC FeSe films on the SrTiO$_3$(001) substrate with similar $\Delta$ can rule out both the interfacial tensile strain \cite{tan2013interface} and ferroelectric phonon screening \cite{xiang2012high} as the primacy for the enhanced superconductivity in terms of the very small lattice mismatch (0.9$\%$) and dramatically different ferroelectric soft phonon frequency in the present case. On the other hand, the interaction between FeSe electrons and SrTiO$_3$ optical phonon at $\sim$ 100 meV that has been proposed to be responsible for raising the superconductivity in SUC FeSe/SrTiO$_3$ films \cite{lee2014interfacial, rademaker2015enhanced, li2015quantum} seems to work here: a nearly identical optical phonon does exist in atatase TiO$_2$ as well \cite{Moser2013tunable}.

\begin{figure}[t]
\includegraphics[width=0.75\columnwidth]{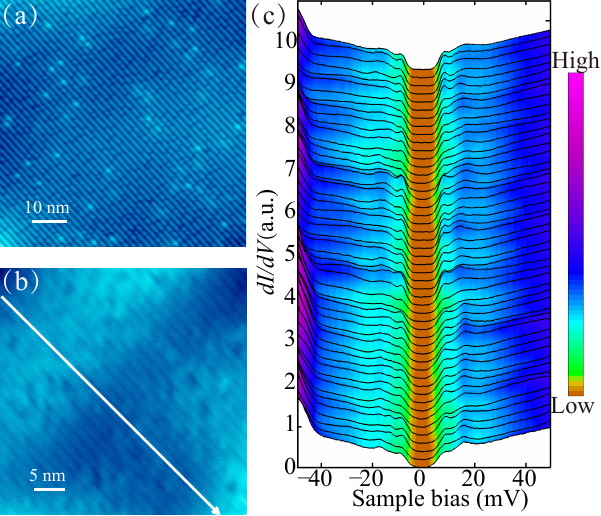}
\caption{(color online) (a) STM topography (70 nm $\times$ 70 nm, $V_\textrm{s}$ = 1.0 V, $I_\textrm{t}$ = 0.03 nA) of anatase TiO$_2$(001) with significantly lower density of oxygen vacancies after post-growth annealing at 750$^\textrm{o}$C. (b) Atomically resolved STM topography (40 nm $\times$ 40 nm, $V_\textrm{s}$ = -50 mV, $I_\textrm{t}$ = 0.1 nA) of SUC FeSe film prepared on the annealed anatase TiO$_2$(001). (c) A series of \textit{dI/dV} spectra (set point: $V_\textrm{s}$ = 50 mV, $I_\textrm{t}$ = 0.4 nA) acquired along the white arrow in b for every nanometer, revealing the almost identical superconducting gap magnitude ($\sim$ 17 meV). The spectra have been vertically shifted relative to each other by 0.2 for clarity.
}
\end{figure}

We now turn to study the charge transfer in the two heterostructures. It has been long believed that they are oxygen vacancies in the SrTiO$_3$ substrate that act as the source of electrons for the overlayer FeSe films \cite{bang2013atomic, zhou2015observation, zhang2015observation}. However, it has not been vividly verified. Although the preparation of SrTiO$_3$ substrate and thus the density of oxygen vacancies might alter substantially among the samples from various groups, a direct relationship between the enhanced superconductivity and oxygen vacancies has never been established \cite{qing2012interface, liu2012electronic, he2013phase, tan2013interface, lee2014interfacial}. To resolve this problem, we have employed STM to image oxygen vacancies and investigate their influence on superconductivity. As shown in Fig.\ 1(c), we can calculate easily the oxygen vacancy density to be $\sim$ 4.6 $\times$ 10$^{-2}$ per nm$^2$, which even remains unchanged after the growth of FeSe films. This amount of vacancies (contributing $\sim$ 0.013 electrons per Fe) is too small by nearly one order of magnitude to account for the necessary doping level of $\sim$ 0.12 electrons per Fe for the occurrence of high-$T_\textrm{c}$ superconductivity in SUC FeSe films \cite{he2013phase}, provided that an oxygen vacancy dopes two electrons into FeSe films. It thus suggests that the oxygen vacancies in TiO$_2$ may not be a sole source for electron doping in SUC FeSe films.

To understand this further, we have annealed the as-grown TiO$_2$/SrTiO$_3$ samples at 750$^\textrm{o}$C to tune the density of oxygen vacancies and investigate their influence on the superconductivity. We find that a significant amount of oxygen atoms that desorb from SrTiO$_3$ can oxidize the overlayer TiO$_2$ islands. As shown in Fig.\ 3(a), it leads to an overall decrease in the density of oxygen vacancies to a level of 6.1 $\times$ 10$^{-3}$ per nm$^2$. However, on the TiO$_2$(001) islands with significantly reduced surface oxygen vacancies (by nearly one order of magnitude), we observe no discernible change in both FeSe morphology [Fig.\ 3(b)] and superconducting gap magnitude $\Delta$ [Fig.\ 3(c)]. The results indicate that the surface oxygen vacancies in TiO$_2$(001) may not be responsible mainly for the charge transfer and thus the high-$T_\textrm{c}$ superconductivity in SUC FeSe films.

Other scenarios, such as interface element diffusion and band bending/alignment across the interface, may be responsible for the electron transfer between FeSe and anatase TiO$_2$. According to Figs.\ 1(e) and 3(b), however, the FeSe/TiO$_2$ interface is abrupt and there is no trace of substitution in FeSe films, indicating that the effect of interface diffusion is not significant. The work function of FeSe was recently measured to be larger than that of anatase TiO$_2$ \cite{privatecommunication}. Upon contact between FeSe and TiO$_2$, the band alignment would lead to an electron transfer from TiO$_2$ to FeSe, consistent with our observation. As thus, the charge transfer mechanism in SUC FeSe/TiO$_2$ might be caused by band alignment, which needs further confirmation in the future. Nevertheless, our direct comparison between TiO$_2$ and SrTiO$_3$ in this study further stresses the important role of interfacial electron-phonon coupling in the enhanced superconductivity in the two heterostructures.

\begin{acknowledgments}
This work is supported by the National Natural Science Foundation of China under Grant No.\ 10721404, 11134008 and 11504196, and the National Basic Research Program of China under Grant No.\ 2009CB929400.  C. L. S. acknowledges support from the Tsinghua University Initiative Scientific Research Program.
\end{acknowledgments}


\begin{thebibliography}{33}%
\makeatletter
\providecommand \@ifxundefined [1]{%
 \@ifx{#1\undefined}
}%
\providecommand \@ifnum [1]{%
 \ifnum #1\expandafter \@firstoftwo
 \else \expandafter \@secondoftwo
 \fi
}%
\providecommand \@ifx [1]{%
 \ifx #1\expandafter \@firstoftwo
 \else \expandafter \@secondoftwo
 \fi
}%
\providecommand \natexlab [1]{#1}%
\providecommand \enquote  [1]{``#1''}%
\providecommand \bibnamefont  [1]{#1}%
\providecommand \bibfnamefont [1]{#1}%
\providecommand \citenamefont [1]{#1}%
\providecommand \href@noop [0]{\@secondoftwo}%
\providecommand \href [0]{\begingroup \@sanitize@url \@href}%
\providecommand \@href[1]{\@@startlink{#1}\@@href}%
\providecommand \@@href[1]{\endgroup#1\@@endlink}%
\providecommand \@sanitize@url [0]{\catcode `\\12\catcode `\$12\catcode
  `\&12\catcode `\#12\catcode `\^12\catcode `\_12\catcode `\%12\relax}%
\providecommand \@@startlink[1]{}%
\providecommand \@@endlink[0]{}%
\providecommand \url  [0]{\begingroup\@sanitize@url \@url }%
\providecommand \@url [1]{\endgroup\@href {#1}{\urlprefix }}%
\providecommand \urlprefix  [0]{URL }%
\providecommand \Eprint [0]{\href }%
\providecommand \doibase [0]{http://dx.doi.org/}%
\providecommand \selectlanguage [0]{\@gobble}%
\providecommand \bibinfo  [0]{\@secondoftwo}%
\providecommand \bibfield  [0]{\@secondoftwo}%
\providecommand \translation [1]{[#1]}%
\providecommand \BibitemOpen [0]{}%
\providecommand \bibitemStop [0]{}%
\providecommand \bibitemNoStop [0]{.\EOS\space}%
\providecommand \EOS [0]{\spacefactor3000\relax}%
\providecommand \BibitemShut  [1]{\csname bibitem#1\endcsname}%
\let\auto@bib@innerbib\@empty
\bibitem [{\citenamefont {Wang}\ \emph {et~al.}(2012)\citenamefont {Wang},
  \citenamefont {Li}, \citenamefont {Zhang}, \citenamefont {Zhang},
  \citenamefont {Zhang}, \citenamefont {Li}, \citenamefont {Ding},
  \citenamefont {Ou}, \citenamefont {Deng}, \citenamefont {Cahng},
  \citenamefont {Wen}, \citenamefont {Song}, \citenamefont {He}, \citenamefont
  {Jia}, \citenamefont {Ji}, \citenamefont {Wang}, \citenamefont {Wang},
  \citenamefont {Chen}, \citenamefont {Ma},\ and\ \citenamefont
  {Xue}}]{qing2012interface}%
  \BibitemOpen
  \bibfield  {author} {\bibinfo {author} {\bibfnamefont {Q.~Y.}\ \bibnamefont
  {Wang}}, \bibinfo {author} {\bibfnamefont {Z.}~\bibnamefont {Li}}, \bibinfo
  {author} {\bibfnamefont {W.~H.}\ \bibnamefont {Zhang}}, \bibinfo {author}
  {\bibfnamefont {Z.~C.}\ \bibnamefont {Zhang}}, \bibinfo {author}
  {\bibfnamefont {J.~S.}\ \bibnamefont {Zhang}}, \bibinfo {author}
  {\bibfnamefont {W.}~\bibnamefont {Li}}, \bibinfo {author} {\bibfnamefont
  {H.}~\bibnamefont {Ding}}, \bibinfo {author} {\bibfnamefont {Y.~B.}\
  \bibnamefont {Ou}}, \bibinfo {author} {\bibfnamefont {P.}~\bibnamefont
  {Deng}}, \bibinfo {author} {\bibfnamefont {K.}~\bibnamefont {Cahng}},
  \bibinfo {author} {\bibfnamefont {J.}~\bibnamefont {Wen}}, \bibinfo {author}
  {\bibfnamefont {C.~L.}\ \bibnamefont {Song}}, \bibinfo {author}
  {\bibfnamefont {K.}~\bibnamefont {He}}, \bibinfo {author} {\bibfnamefont
  {J.~F.}\ \bibnamefont {Jia}}, \bibinfo {author} {\bibfnamefont {S.~H.}\
  \bibnamefont {Ji}}, \bibinfo {author} {\bibfnamefont {Y.~Y.}\ \bibnamefont
  {Wang}}, \bibinfo {author} {\bibfnamefont {L.~L.}\ \bibnamefont {Wang}},
  \bibinfo {author} {\bibfnamefont {X.}~\bibnamefont {Chen}}, \bibinfo {author}
  {\bibfnamefont {X.~C.}\ \bibnamefont {Ma}}, \ and\ \bibinfo {author}
  {\bibfnamefont {Q.~K.}\ \bibnamefont {Xue}},\ }\href {\doibase
  10.1088/0256-307X/29/3/037402} {\bibfield  {journal} {\bibinfo  {journal}
  {Chin. Phys. Lett.}\ }\textbf {\bibinfo {volume} {29}},\ \bibinfo {pages}
  {037402} (\bibinfo {year} {2012})}\BibitemShut {NoStop}%
\bibitem [{\citenamefont {Liu}\ \emph {et~al.}(2012)\citenamefont {Liu},
  \citenamefont {Zhang}, \citenamefont {Mou}, \citenamefont {He}, \citenamefont
  {Ou}, \citenamefont {Wang}, \citenamefont {Li}, \citenamefont {Wang},
  \citenamefont {Zhao}, \citenamefont {He}, \citenamefont {Peng}, \citenamefont
  {Liu}, \citenamefont {Chen}, \citenamefont {Yu}, \citenamefont {Liu},
  \citenamefont {Dong}, \citenamefont {Zhang}, \citenamefont {Chen},
  \citenamefont {Xue}, \citenamefont {Hu}, \citenamefont {Chen}, \citenamefont
  {Ma}, \citenamefont {Xue},\ and\ \citenamefont {Zhou}}]{liu2012electronic}%
  \BibitemOpen
  \bibfield  {author} {\bibinfo {author} {\bibfnamefont {D.~F.}\ \bibnamefont
  {Liu}}, \bibinfo {author} {\bibfnamefont {W.~H.}\ \bibnamefont {Zhang}},
  \bibinfo {author} {\bibfnamefont {D.~X.}\ \bibnamefont {Mou}}, \bibinfo
  {author} {\bibfnamefont {J.~F.}\ \bibnamefont {He}}, \bibinfo {author}
  {\bibfnamefont {Y.~B.}\ \bibnamefont {Ou}}, \bibinfo {author} {\bibfnamefont
  {Q.~Y.}\ \bibnamefont {Wang}}, \bibinfo {author} {\bibfnamefont
  {Z.}~\bibnamefont {Li}}, \bibinfo {author} {\bibfnamefont {L.~L.}\
  \bibnamefont {Wang}}, \bibinfo {author} {\bibfnamefont {L.}~\bibnamefont
  {Zhao}}, \bibinfo {author} {\bibfnamefont {S.~L.}\ \bibnamefont {He}},
  \bibinfo {author} {\bibfnamefont {Y.~Y.}\ \bibnamefont {Peng}}, \bibinfo
  {author} {\bibfnamefont {X.}~\bibnamefont {Liu}}, \bibinfo {author}
  {\bibfnamefont {X.~Y.}\ \bibnamefont {Chen}}, \bibinfo {author}
  {\bibfnamefont {L.}~\bibnamefont {Yu}}, \bibinfo {author} {\bibfnamefont
  {G.~D.}\ \bibnamefont {Liu}}, \bibinfo {author} {\bibfnamefont {X.~L.}\
  \bibnamefont {Dong}}, \bibinfo {author} {\bibfnamefont {J.}~\bibnamefont
  {Zhang}}, \bibinfo {author} {\bibfnamefont {C.~T.}\ \bibnamefont {Chen}},
  \bibinfo {author} {\bibfnamefont {Z.~Y.}\ \bibnamefont {Xue}}, \bibinfo
  {author} {\bibfnamefont {J.~P.}\ \bibnamefont {Hu}}, \bibinfo {author}
  {\bibfnamefont {X.}~\bibnamefont {Chen}}, \bibinfo {author} {\bibfnamefont
  {X.~C.}\ \bibnamefont {Ma}}, \bibinfo {author} {\bibfnamefont {Q.~K.}\
  \bibnamefont {Xue}}, \ and\ \bibinfo {author} {\bibfnamefont {X.~J.}\
  \bibnamefont {Zhou}},\ }\href {\doibase 10.1038/ncomms1946} {\bibfield
  {journal} {\bibinfo  {journal} {Nat. Commun.}\ }\textbf {\bibinfo {volume}
  {3}},\ \bibinfo {pages} {931} (\bibinfo {year} {2012})}\BibitemShut {NoStop}%
\bibitem [{\citenamefont {He}\ \emph {et~al.}(2013)\citenamefont {He},
  \citenamefont {He}, \citenamefont {Zhang}, \citenamefont {Zhao},
  \citenamefont {Liu}, \citenamefont {Liu}, \citenamefont {Mou}, \citenamefont
  {Ou}, \citenamefont {Wang}, \citenamefont {Li}, \citenamefont {Wang},
  \citenamefont {Peng}, \citenamefont {Liu}, \citenamefont {Chen},
  \citenamefont {Yu}, \citenamefont {Liu}, \citenamefont {Dong}, \citenamefont
  {Zhang}, \citenamefont {Chen}, \citenamefont {Xu}, \citenamefont {Chen},
  \citenamefont {Ma}, \citenamefont {Xue},\ and\ \citenamefont
  {Zhou}}]{he2013phase}%
  \BibitemOpen
  \bibfield  {author} {\bibinfo {author} {\bibfnamefont {S.~L.}\ \bibnamefont
  {He}}, \bibinfo {author} {\bibfnamefont {J.~F.}\ \bibnamefont {He}}, \bibinfo
  {author} {\bibfnamefont {W.~H.}\ \bibnamefont {Zhang}}, \bibinfo {author}
  {\bibfnamefont {L.}~\bibnamefont {Zhao}}, \bibinfo {author} {\bibfnamefont
  {D.~F.}\ \bibnamefont {Liu}}, \bibinfo {author} {\bibfnamefont
  {X.}~\bibnamefont {Liu}}, \bibinfo {author} {\bibfnamefont {D.~X.}\
  \bibnamefont {Mou}}, \bibinfo {author} {\bibfnamefont {Y.~B.}\ \bibnamefont
  {Ou}}, \bibinfo {author} {\bibfnamefont {Q.~Y.}\ \bibnamefont {Wang}},
  \bibinfo {author} {\bibfnamefont {Z.}~\bibnamefont {Li}}, \bibinfo {author}
  {\bibfnamefont {L.~L.}\ \bibnamefont {Wang}}, \bibinfo {author}
  {\bibfnamefont {Y.~Y.}\ \bibnamefont {Peng}}, \bibinfo {author}
  {\bibfnamefont {Y.}~\bibnamefont {Liu}}, \bibinfo {author} {\bibfnamefont
  {C.~Y.}\ \bibnamefont {Chen}}, \bibinfo {author} {\bibfnamefont
  {L.}~\bibnamefont {Yu}}, \bibinfo {author} {\bibfnamefont {G.~D.}\
  \bibnamefont {Liu}}, \bibinfo {author} {\bibfnamefont {X.~L.}\ \bibnamefont
  {Dong}}, \bibinfo {author} {\bibfnamefont {J.}~\bibnamefont {Zhang}},
  \bibinfo {author} {\bibfnamefont {C.~T.}\ \bibnamefont {Chen}}, \bibinfo
  {author} {\bibfnamefont {Z.~Y.}\ \bibnamefont {Xu}}, \bibinfo {author}
  {\bibfnamefont {X.}~\bibnamefont {Chen}}, \bibinfo {author} {\bibfnamefont
  {X.~C.}\ \bibnamefont {Ma}}, \bibinfo {author} {\bibfnamefont {Q.~K.}\
  \bibnamefont {Xue}}, \ and\ \bibinfo {author} {\bibfnamefont {X.~J.}\
  \bibnamefont {Zhou}},\ }\href {\doibase 10.1038/nmat3648} {\bibfield
  {journal} {\bibinfo  {journal} {Nat. Mater.}\ }\textbf {\bibinfo {volume}
  {12}},\ \bibinfo {pages} {605} (\bibinfo {year} {2013})}\BibitemShut
  {NoStop}%
\bibitem [{\citenamefont {Tan}\ \emph {et~al.}(2013)\citenamefont {Tan},
  \citenamefont {Zhang}, \citenamefont {Xia}, \citenamefont {Ye}, \citenamefont
  {Chen}, \citenamefont {Xie}, \citenamefont {Peng}, \citenamefont {Xu},
  \citenamefont {Fan}, \citenamefont {Xu}, \citenamefont {Jiang}, \citenamefont
  {Zhang}, \citenamefont {Lai}, \citenamefont {Xiang}, \citenamefont {Hu},
  \citenamefont {Xie},\ and\ \citenamefont {Feng}}]{tan2013interface}%
  \BibitemOpen
  \bibfield  {author} {\bibinfo {author} {\bibfnamefont {S.~Y.}\ \bibnamefont
  {Tan}}, \bibinfo {author} {\bibfnamefont {Y.}~\bibnamefont {Zhang}}, \bibinfo
  {author} {\bibfnamefont {M.}~\bibnamefont {Xia}}, \bibinfo {author}
  {\bibfnamefont {Z.~R.}\ \bibnamefont {Ye}}, \bibinfo {author} {\bibfnamefont
  {F.}~\bibnamefont {Chen}}, \bibinfo {author} {\bibfnamefont {X.}~\bibnamefont
  {Xie}}, \bibinfo {author} {\bibfnamefont {R.}~\bibnamefont {Peng}}, \bibinfo
  {author} {\bibfnamefont {D.~F.}\ \bibnamefont {Xu}}, \bibinfo {author}
  {\bibfnamefont {Q.}~\bibnamefont {Fan}}, \bibinfo {author} {\bibfnamefont
  {H.~C.}\ \bibnamefont {Xu}}, \bibinfo {author} {\bibfnamefont
  {J.}~\bibnamefont {Jiang}}, \bibinfo {author} {\bibfnamefont
  {T.}~\bibnamefont {Zhang}}, \bibinfo {author} {\bibfnamefont {X.~C.}\
  \bibnamefont {Lai}}, \bibinfo {author} {\bibfnamefont {T.}~\bibnamefont
  {Xiang}}, \bibinfo {author} {\bibfnamefont {J.~P.}\ \bibnamefont {Hu}},
  \bibinfo {author} {\bibfnamefont {N.~P.}\ \bibnamefont {Xie}}, \ and\
  \bibinfo {author} {\bibfnamefont {D.~L.}\ \bibnamefont {Feng}},\ }\href
  {\doibase 10.1038/nmat3654} {\bibfield  {journal} {\bibinfo  {journal} {Nat.
  Mater.}\ }\textbf {\bibinfo {volume} {12}},\ \bibinfo {pages} {634} (\bibinfo
  {year} {2013})}\BibitemShut {NoStop}%
\bibitem [{\citenamefont {Xiang}\ \emph {et~al.}(2012)\citenamefont {Xiang},
  \citenamefont {Wang}, \citenamefont {Wang}, \citenamefont {Wang},\ and\
  \citenamefont {Lee}}]{xiang2012high}%
  \BibitemOpen
  \bibfield  {author} {\bibinfo {author} {\bibfnamefont {Y.~Y.}\ \bibnamefont
  {Xiang}}, \bibinfo {author} {\bibfnamefont {F.}~\bibnamefont {Wang}},
  \bibinfo {author} {\bibfnamefont {D.}~\bibnamefont {Wang}}, \bibinfo {author}
  {\bibfnamefont {Q.~H.}\ \bibnamefont {Wang}}, \ and\ \bibinfo {author}
  {\bibfnamefont {D.~H.}\ \bibnamefont {Lee}},\ }\href {\doibase
  10.1103/PhysRevB.86.134508} {\bibfield  {journal} {\bibinfo  {journal} {Phys.
  Rev. B}\ }\textbf {\bibinfo {volume} {86}},\ \bibinfo {pages} {134508}
  (\bibinfo {year} {2012})}\BibitemShut {NoStop}%
\bibitem [{\citenamefont {Bang}\ \emph {et~al.}(2013)\citenamefont {Bang},
  \citenamefont {Li}, \citenamefont {Sun}, \citenamefont {Samanta},
  \citenamefont {Zhang}, \citenamefont {Zhang}, \citenamefont {Wang},
  \citenamefont {Chen}, \citenamefont {Ma}, \citenamefont {Xue},\ and\
  \citenamefont {Zhang}}]{bang2013atomic}%
  \BibitemOpen
  \bibfield  {author} {\bibinfo {author} {\bibfnamefont {J.}~\bibnamefont
  {Bang}}, \bibinfo {author} {\bibfnamefont {Z.}~\bibnamefont {Li}}, \bibinfo
  {author} {\bibfnamefont {Y.~Y.}\ \bibnamefont {Sun}}, \bibinfo {author}
  {\bibfnamefont {A.}~\bibnamefont {Samanta}}, \bibinfo {author} {\bibfnamefont
  {Y.~Y.}\ \bibnamefont {Zhang}}, \bibinfo {author} {\bibfnamefont {W.~H.}\
  \bibnamefont {Zhang}}, \bibinfo {author} {\bibfnamefont {L.}~\bibnamefont
  {Wang}}, \bibinfo {author} {\bibfnamefont {X.}~\bibnamefont {Chen}}, \bibinfo
  {author} {\bibfnamefont {X.~C.}\ \bibnamefont {Ma}}, \bibinfo {author}
  {\bibfnamefont {Q.~K.}\ \bibnamefont {Xue}}, \ and\ \bibinfo {author}
  {\bibfnamefont {S.~B.}\ \bibnamefont {Zhang}},\ }\href {\doibase
  10.1103/PhysRevB.87.220503} {\bibfield  {journal} {\bibinfo  {journal} {Phys.
  Rev. B}\ }\textbf {\bibinfo {volume} {87}},\ \bibinfo {pages} {220503}
  (\bibinfo {year} {2013})}\BibitemShut {NoStop}%
\bibitem [{\citenamefont {Lee}\ \emph {et~al.}(2014)\citenamefont {Lee},
  \citenamefont {Schmitt}, \citenamefont {Moore}, \citenamefont {Johnston},
  \citenamefont {Cui}, \citenamefont {Li}, \citenamefont {Yi}, \citenamefont
  {Liu}, \citenamefont {Hashimoto}, \citenamefont {Zhang}, \citenamefont {Lu},
  \citenamefont {Devereaux}, \citenamefont {Lee},\ and\ \citenamefont
  {Shen}}]{lee2014interfacial}%
  \BibitemOpen
  \bibfield  {author} {\bibinfo {author} {\bibfnamefont {J.~J.}\ \bibnamefont
  {Lee}}, \bibinfo {author} {\bibfnamefont {F.~T.}\ \bibnamefont {Schmitt}},
  \bibinfo {author} {\bibfnamefont {R.~G.}\ \bibnamefont {Moore}}, \bibinfo
  {author} {\bibfnamefont {S.}~\bibnamefont {Johnston}}, \bibinfo {author}
  {\bibfnamefont {Y.~T.}\ \bibnamefont {Cui}}, \bibinfo {author} {\bibfnamefont
  {W.}~\bibnamefont {Li}}, \bibinfo {author} {\bibfnamefont {M.}~\bibnamefont
  {Yi}}, \bibinfo {author} {\bibfnamefont {Z.~K.}\ \bibnamefont {Liu}},
  \bibinfo {author} {\bibfnamefont {M.}~\bibnamefont {Hashimoto}}, \bibinfo
  {author} {\bibfnamefont {Y.}~\bibnamefont {Zhang}}, \bibinfo {author}
  {\bibfnamefont {D.~H.}\ \bibnamefont {Lu}}, \bibinfo {author} {\bibfnamefont
  {T.~P.}\ \bibnamefont {Devereaux}}, \bibinfo {author} {\bibfnamefont {D.~H.}\
  \bibnamefont {Lee}}, \ and\ \bibinfo {author} {\bibfnamefont {Z.~X.}\
  \bibnamefont {Shen}},\ }\href {\doibase 10.1038/nature13894} {\bibfield
  {journal} {\bibinfo  {journal} {Nature}\ }\textbf {\bibinfo {volume} {515}},\
  \bibinfo {pages} {245} (\bibinfo {year} {2014})}\BibitemShut {NoStop}%
\bibitem [{\citenamefont {Miyata}\ \emph {et~al.}(2015)\citenamefont {Miyata},
  \citenamefont {Nakayama}, \citenamefont {Sugawara}, \citenamefont {Sato},\
  and\ \citenamefont {Takahashi}}]{miyata2015high}%
  \BibitemOpen
  \bibfield  {author} {\bibinfo {author} {\bibfnamefont {Y.}~\bibnamefont
  {Miyata}}, \bibinfo {author} {\bibfnamefont {K.}~\bibnamefont {Nakayama}},
  \bibinfo {author} {\bibfnamefont {K.}~\bibnamefont {Sugawara}}, \bibinfo
  {author} {\bibfnamefont {T.}~\bibnamefont {Sato}}, \ and\ \bibinfo {author}
  {\bibfnamefont {T.}~\bibnamefont {Takahashi}},\ }\href {\doibase
  10.1038/nmat4302} {\bibfield  {journal} {\bibinfo  {journal} {Nat. Mater.}\
  }\textbf {\bibinfo {volume} {14}},\ \bibinfo {pages} {775} (\bibinfo {year}
  {2015})}\BibitemShut {NoStop}%
\bibitem [{\citenamefont {Rademaker}\ \emph {et~al.}(2016)\citenamefont
  {Rademaker}, \citenamefont {Wang}, \citenamefont {Berlijn},\ and\
  \citenamefont {Johnston}}]{rademaker2015enhanced}%
  \BibitemOpen
  \bibfield  {author} {\bibinfo {author} {\bibfnamefont {L.}~\bibnamefont
  {Rademaker}}, \bibinfo {author} {\bibfnamefont {Y.}~\bibnamefont {Wang}},
  \bibinfo {author} {\bibfnamefont {T.}~\bibnamefont {Berlijn}}, \ and\
  \bibinfo {author} {\bibfnamefont {S.}~\bibnamefont {Johnston}},\ }\href
  {\doibase 10.1088/1367-2630/18/2/022001} {\bibfield  {journal} {\bibinfo
  {journal} {arXiv preprint arXiv:1507.03967}\ }\textbf {\bibinfo {volume}
  {18}},\ \bibinfo {pages} {022001} (\bibinfo {year} {2016})}\BibitemShut
  {NoStop}%
\bibitem [{\citenamefont {Li}\ \emph {et~al.}(2015)\citenamefont {Li},
  \citenamefont {Wang}, \citenamefont {Yao},\ and\ \citenamefont
  {Lee}}]{li2015quantum}%
  \BibitemOpen
  \bibfield  {author} {\bibinfo {author} {\bibfnamefont {Z.~X.}\ \bibnamefont
  {Li}}, \bibinfo {author} {\bibfnamefont {F.}~\bibnamefont {Wang}}, \bibinfo
  {author} {\bibfnamefont {H.}~\bibnamefont {Yao}}, \ and\ \bibinfo {author}
  {\bibfnamefont {D.~H.}\ \bibnamefont {Lee}},\ }\href@noop {} {\bibfield
  {journal} {\bibinfo  {journal} {arXiv preprint arXiv:1512.06179}\ } (\bibinfo
  {year} {2015})}\BibitemShut {NoStop}%
\bibitem [{\citenamefont {Nakayama}\ \emph {et~al.}(2014)\citenamefont
  {Nakayama}, \citenamefont {Miyata}, \citenamefont {Phan}, \citenamefont
  {Sato}, \citenamefont {Tanabe}, \citenamefont {Urata}, \citenamefont
  {Tanigaki},\ and\ \citenamefont {Takahashi}}]{Nakayama2014reconstruction}%
  \BibitemOpen
  \bibfield  {author} {\bibinfo {author} {\bibfnamefont {K.}~\bibnamefont
  {Nakayama}}, \bibinfo {author} {\bibfnamefont {Y.}~\bibnamefont {Miyata}},
  \bibinfo {author} {\bibfnamefont {G.}~\bibnamefont {Phan}}, \bibinfo {author}
  {\bibfnamefont {T.}~\bibnamefont {Sato}}, \bibinfo {author} {\bibfnamefont
  {Y.}~\bibnamefont {Tanabe}}, \bibinfo {author} {\bibfnamefont
  {T.}~\bibnamefont {Urata}}, \bibinfo {author} {\bibfnamefont
  {K.}~\bibnamefont {Tanigaki}}, \ and\ \bibinfo {author} {\bibfnamefont
  {T.}~\bibnamefont {Takahashi}},\ }\href {\doibase
  10.1103/PhysRevLett.113.237001} {\bibfield  {journal} {\bibinfo  {journal}
  {Phys. Rev. Lett.}\ }\textbf {\bibinfo {volume} {113}},\ \bibinfo {pages}
  {237001} (\bibinfo {year} {2014})}\BibitemShut {NoStop}%
\bibitem [{\citenamefont {Song}\ \emph {et~al.}(2011)\citenamefont {Song},
  \citenamefont {Wang}, \citenamefont {Cheng}, \citenamefont {Jiang},
  \citenamefont {Li}, \citenamefont {Zhang}, \citenamefont {Li}, \citenamefont
  {He}, \citenamefont {Wang}, \citenamefont {Jia}, \citenamefont {Hung},
  \citenamefont {Wu}, \citenamefont {Ma}, \citenamefont {Chen},\ and\
  \citenamefont {Xue}}]{song2011direct}%
  \BibitemOpen
  \bibfield  {author} {\bibinfo {author} {\bibfnamefont {C.~L.}\ \bibnamefont
  {Song}}, \bibinfo {author} {\bibfnamefont {Y.~L.}\ \bibnamefont {Wang}},
  \bibinfo {author} {\bibfnamefont {P.}~\bibnamefont {Cheng}}, \bibinfo
  {author} {\bibfnamefont {Y.~P.}\ \bibnamefont {Jiang}}, \bibinfo {author}
  {\bibfnamefont {W.}~\bibnamefont {Li}}, \bibinfo {author} {\bibfnamefont
  {T.}~\bibnamefont {Zhang}}, \bibinfo {author} {\bibfnamefont
  {Z.}~\bibnamefont {Li}}, \bibinfo {author} {\bibfnamefont {K.}~\bibnamefont
  {He}}, \bibinfo {author} {\bibfnamefont {L.}~\bibnamefont {Wang}}, \bibinfo
  {author} {\bibfnamefont {J.~F.}\ \bibnamefont {Jia}}, \bibinfo {author}
  {\bibfnamefont {H.-H.}\ \bibnamefont {Hung}}, \bibinfo {author}
  {\bibfnamefont {C.~J.}\ \bibnamefont {Wu}}, \bibinfo {author} {\bibfnamefont
  {X.~C.}\ \bibnamefont {Ma}}, \bibinfo {author} {\bibfnamefont
  {X.}~\bibnamefont {Chen}}, \ and\ \bibinfo {author} {\bibfnamefont {Q.~K.}\
  \bibnamefont {Xue}},\ }\href {\doibase 10.1126/science.1202226} {\bibfield
  {journal} {\bibinfo  {journal} {Science}\ }\textbf {\bibinfo {volume}
  {332}},\ \bibinfo {pages} {1410} (\bibinfo {year} {2011})}\BibitemShut
  {NoStop}%
\bibitem [{\citenamefont {Peng}\ \emph {et~al.}(2014)\citenamefont {Peng},
  \citenamefont {Xu}, \citenamefont {Tan}, \citenamefont {Cao}, \citenamefont
  {Xia}, \citenamefont {Shen}, \citenamefont {Huang}, \citenamefont {Wen},
  \citenamefont {Song}, \citenamefont {Zhang}, \citenamefont {Zie},\ and\
  \citenamefont {Feng}}]{peng2014tuning}%
  \BibitemOpen
  \bibfield  {author} {\bibinfo {author} {\bibfnamefont {R.}~\bibnamefont
  {Peng}}, \bibinfo {author} {\bibfnamefont {H.~C.}\ \bibnamefont {Xu}},
  \bibinfo {author} {\bibfnamefont {S.~Y.}\ \bibnamefont {Tan}}, \bibinfo
  {author} {\bibfnamefont {H.~Y.}\ \bibnamefont {Cao}}, \bibinfo {author}
  {\bibfnamefont {M.}~\bibnamefont {Xia}}, \bibinfo {author} {\bibfnamefont
  {X.~P.}\ \bibnamefont {Shen}}, \bibinfo {author} {\bibfnamefont {Z.~C.}\
  \bibnamefont {Huang}}, \bibinfo {author} {\bibfnamefont {C.~H.~P.}\
  \bibnamefont {Wen}}, \bibinfo {author} {\bibfnamefont {Q.}~\bibnamefont
  {Song}}, \bibinfo {author} {\bibfnamefont {T.}~\bibnamefont {Zhang}},
  \bibinfo {author} {\bibfnamefont {B.~P.}\ \bibnamefont {Zie}}, \ and\
  \bibinfo {author} {\bibfnamefont {D.~L.}\ \bibnamefont {Feng}},\ }\href
  {\doibase 10.1038/ncomms6044} {\bibfield  {journal} {\bibinfo  {journal}
  {Nat. Commun.}\ }\textbf {\bibinfo {volume} {5}},\ \bibinfo {pages} {5044}
  (\bibinfo {year} {2014})}\BibitemShut {NoStop}%
\bibitem [{\citenamefont {Zhou}\ \emph {et~al.}(2016)\citenamefont {Zhou},
  \citenamefont {Zhang}, \citenamefont {Liu}, \citenamefont {Tang},
  \citenamefont {Wang}, \citenamefont {Li}, \citenamefont {Song}, \citenamefont
  {Ji}, \citenamefont {He}, \citenamefont {Wang}, \citenamefont {Ma},\ and\
  \citenamefont {Xue}}]{zhou2015observation}%
  \BibitemOpen
  \bibfield  {author} {\bibinfo {author} {\bibfnamefont {G.~Y.}\ \bibnamefont
  {Zhou}}, \bibinfo {author} {\bibfnamefont {D.}~\bibnamefont {Zhang}},
  \bibinfo {author} {\bibfnamefont {C.}~\bibnamefont {Liu}}, \bibinfo {author}
  {\bibfnamefont {C.~J.}\ \bibnamefont {Tang}}, \bibinfo {author}
  {\bibfnamefont {X.}~\bibnamefont {Wang}}, \bibinfo {author} {\bibfnamefont
  {Z.}~\bibnamefont {Li}}, \bibinfo {author} {\bibfnamefont {C.}~\bibnamefont
  {Song}}, \bibinfo {author} {\bibfnamefont {S.}~\bibnamefont {Ji}}, \bibinfo
  {author} {\bibfnamefont {K.}~\bibnamefont {He}}, \bibinfo {author}
  {\bibfnamefont {L.}~\bibnamefont {Wang}}, \bibinfo {author} {\bibfnamefont
  {X.~C.}\ \bibnamefont {Ma}}, \ and\ \bibinfo {author} {\bibfnamefont {Q.~K.}\
  \bibnamefont {Xue}},\ }\href {\doibase 10.1063/1.4950964} {\bibfield
  {journal} {\bibinfo  {journal} {Appl. Phys. Lett.}\ }\textbf {\bibinfo
  {volume} {108}},\ \bibinfo {pages} {202603} (\bibinfo {year}
  {2016})}\BibitemShut {NoStop}%
\bibitem [{\citenamefont {Zhang}\ \emph
  {et~al.}(2015{\natexlab{a}})\citenamefont {Zhang}, \citenamefont {Peng},
  \citenamefont {Qian}, \citenamefont {Richard}, \citenamefont {Shi},
  \citenamefont {Ma}, \citenamefont {Fu}, \citenamefont {Guo}, \citenamefont
  {Han}, \citenamefont {Wang}, \citenamefont {Wang}, \citenamefont {Xue},
  \citenamefont {Hu}, \citenamefont {Sun},\ and\ \citenamefont
  {Ding}}]{zhang2015observation}%
  \BibitemOpen
  \bibfield  {author} {\bibinfo {author} {\bibfnamefont {P.}~\bibnamefont
  {Zhang}}, \bibinfo {author} {\bibfnamefont {X.~L.}\ \bibnamefont {Peng}},
  \bibinfo {author} {\bibfnamefont {T.}~\bibnamefont {Qian}}, \bibinfo {author}
  {\bibfnamefont {P.}~\bibnamefont {Richard}}, \bibinfo {author} {\bibfnamefont
  {X.}~\bibnamefont {Shi}}, \bibinfo {author} {\bibfnamefont {J.-Z.}\
  \bibnamefont {Ma}}, \bibinfo {author} {\bibfnamefont {B.-B.}\ \bibnamefont
  {Fu}}, \bibinfo {author} {\bibfnamefont {Y.~L.}\ \bibnamefont {Guo}},
  \bibinfo {author} {\bibfnamefont {Z.~Q.}\ \bibnamefont {Han}}, \bibinfo
  {author} {\bibfnamefont {S.~C.}\ \bibnamefont {Wang}}, \bibinfo {author}
  {\bibfnamefont {L.~L.}\ \bibnamefont {Wang}}, \bibinfo {author}
  {\bibfnamefont {Q.~K.}\ \bibnamefont {Xue}}, \bibinfo {author} {\bibfnamefont
  {J.~P.}\ \bibnamefont {Hu}}, \bibinfo {author} {\bibfnamefont {Y.~J.}\
  \bibnamefont {Sun}}, \ and\ \bibinfo {author} {\bibfnamefont
  {H.}~\bibnamefont {Ding}},\ }\href@noop {} {\bibfield  {journal} {\bibinfo
  {journal} {arXiv preprint arXiv:1512.01949}\ } (\bibinfo {year}
  {2015}{\natexlab{a}})}\BibitemShut {NoStop}%
\bibitem [{\citenamefont {Burrard-Lucas}\ \emph {et~al.}(2013)\citenamefont
  {Burrard-Lucas}, \citenamefont {Free}, \citenamefont {Sedlmaier},
  \citenamefont {Wright}, \citenamefont {Cassidy}, \citenamefont {Hara},
  \citenamefont {Corkett}, \citenamefont {Lancaster}, \citenamefont {Baker},
  \citenamefont {Blundell},\ and\ \citenamefont
  {Clarke}}]{burrard2013enhancement}%
  \BibitemOpen
  \bibfield  {author} {\bibinfo {author} {\bibfnamefont {M.}~\bibnamefont
  {Burrard-Lucas}}, \bibinfo {author} {\bibfnamefont {D.~G.}\ \bibnamefont
  {Free}}, \bibinfo {author} {\bibfnamefont {S.~J.}\ \bibnamefont {Sedlmaier}},
  \bibinfo {author} {\bibfnamefont {J.~D.}\ \bibnamefont {Wright}}, \bibinfo
  {author} {\bibfnamefont {S.~J.}\ \bibnamefont {Cassidy}}, \bibinfo {author}
  {\bibfnamefont {Y.}~\bibnamefont {Hara}}, \bibinfo {author} {\bibfnamefont
  {A.~J.}\ \bibnamefont {Corkett}}, \bibinfo {author} {\bibfnamefont
  {T.}~\bibnamefont {Lancaster}}, \bibinfo {author} {\bibfnamefont {P.~J.}\
  \bibnamefont {Baker}}, \bibinfo {author} {\bibfnamefont {S.~J.}\ \bibnamefont
  {Blundell}}, \ and\ \bibinfo {author} {\bibfnamefont {S.~J.}\ \bibnamefont
  {Clarke}},\ }\href {\doibase 10.1038/nmat3464} {\bibfield  {journal}
  {\bibinfo  {journal} {Nat. Mater.}\ }\textbf {\bibinfo {volume} {12}},\
  \bibinfo {pages} {15} (\bibinfo {year} {2013})}\BibitemShut {NoStop}%
\bibitem [{\citenamefont {Lu}\ \emph {et~al.}(2015)\citenamefont {Lu},
  \citenamefont {Wang}, \citenamefont {Wu}, \citenamefont {Wu}, \citenamefont
  {Zhao}, \citenamefont {Zeng}, \citenamefont {Luo}, \citenamefont {Wu},
  \citenamefont {Bao}, \citenamefont {Zhang},\ and\ \citenamefont
  {Chen}}]{lu2014coexistence}%
  \BibitemOpen
  \bibfield  {author} {\bibinfo {author} {\bibfnamefont {X.~F.}\ \bibnamefont
  {Lu}}, \bibinfo {author} {\bibfnamefont {N.~Z.}\ \bibnamefont {Wang}},
  \bibinfo {author} {\bibfnamefont {H.}~\bibnamefont {Wu}}, \bibinfo {author}
  {\bibfnamefont {Y.~P.}\ \bibnamefont {Wu}}, \bibinfo {author} {\bibfnamefont
  {D.}~\bibnamefont {Zhao}}, \bibinfo {author} {\bibfnamefont {X.~Z.}\
  \bibnamefont {Zeng}}, \bibinfo {author} {\bibfnamefont {X.~G.}\ \bibnamefont
  {Luo}}, \bibinfo {author} {\bibfnamefont {T.}~\bibnamefont {Wu}}, \bibinfo
  {author} {\bibfnamefont {W.}~\bibnamefont {Bao}}, \bibinfo {author}
  {\bibfnamefont {G.~H.}\ \bibnamefont {Zhang}}, \ and\ \bibinfo {author}
  {\bibfnamefont {X.~H.}\ \bibnamefont {Chen}},\ }\href {\doibase
  10.1038/nmat4155} {\bibfield  {journal} {\bibinfo  {journal} {Nat. Mater.}\
  }\textbf {\bibinfo {volume} {14}},\ \bibinfo {pages} {325} (\bibinfo {year}
  {2015})}\BibitemShut {NoStop}%
\bibitem [{\citenamefont {Lei}\ \emph {et~al.}(2016)\citenamefont {Lei},
  \citenamefont {Cui}, \citenamefont {Xiang}, \citenamefont {Shang},
  \citenamefont {Wang}, \citenamefont {Ye}, \citenamefont {Luo}, \citenamefont
  {Wu}, \citenamefont {Sun},\ and\ \citenamefont {Chen}}]{lei2015evolution}%
  \BibitemOpen
  \bibfield  {author} {\bibinfo {author} {\bibfnamefont {B.}~\bibnamefont
  {Lei}}, \bibinfo {author} {\bibfnamefont {J.~H.}\ \bibnamefont {Cui}},
  \bibinfo {author} {\bibfnamefont {Z.~J.}\ \bibnamefont {Xiang}}, \bibinfo
  {author} {\bibfnamefont {C.}~\bibnamefont {Shang}}, \bibinfo {author}
  {\bibfnamefont {N.~Z.}\ \bibnamefont {Wang}}, \bibinfo {author}
  {\bibfnamefont {G.~J.}\ \bibnamefont {Ye}}, \bibinfo {author} {\bibfnamefont
  {X.~G.}\ \bibnamefont {Luo}}, \bibinfo {author} {\bibfnamefont
  {T.}~\bibnamefont {Wu}}, \bibinfo {author} {\bibfnamefont {Z.}~\bibnamefont
  {Sun}}, \ and\ \bibinfo {author} {\bibfnamefont {X.~H.}\ \bibnamefont
  {Chen}},\ }\href {\doibase 10.1103/PhysRevLett.116.077002} {\bibfield
  {journal} {\bibinfo  {journal} {Phys. Rev. Lett.}\ }\textbf {\bibinfo
  {volume} {116}},\ \bibinfo {pages} {077002} (\bibinfo {year}
  {2016})}\BibitemShut {NoStop}%
\bibitem [{\citenamefont {Song}\ \emph {et~al.}(2016)\citenamefont {Song},
  \citenamefont {Zhang}, \citenamefont {Zhong}, \citenamefont {Hu},
  \citenamefont {Ji}, \citenamefont {Wang}, \citenamefont {He}, \citenamefont
  {Ma},\ and\ \citenamefont {Xue}}]{Song2016observation}%
  \BibitemOpen
  \bibfield  {author} {\bibinfo {author} {\bibfnamefont {C.~L.}\ \bibnamefont
  {Song}}, \bibinfo {author} {\bibfnamefont {H.~M.}\ \bibnamefont {Zhang}},
  \bibinfo {author} {\bibfnamefont {Y.}~\bibnamefont {Zhong}}, \bibinfo
  {author} {\bibfnamefont {X.~P.}\ \bibnamefont {Hu}}, \bibinfo {author}
  {\bibfnamefont {S.~H.}\ \bibnamefont {Ji}}, \bibinfo {author} {\bibfnamefont
  {L.}~\bibnamefont {Wang}}, \bibinfo {author} {\bibfnamefont {K.}~\bibnamefont
  {He}}, \bibinfo {author} {\bibfnamefont {X.~C.}\ \bibnamefont {Ma}}, \ and\
  \bibinfo {author} {\bibfnamefont {Q.~K.}\ \bibnamefont {Xue}},\ }\href
  {\doibase 10.1103/PhysRevLett.116.157001} {\bibfield  {journal} {\bibinfo
  {journal} {Phys. Rev. Lett.}\ }\textbf {\bibinfo {volume} {116}},\ \bibinfo
  {pages} {157001} (\bibinfo {year} {2016})}\BibitemShut {NoStop}%
\bibitem [{\citenamefont {Zhang}\ \emph
  {et~al.}(2015{\natexlab{b}})\citenamefont {Zhang}, \citenamefont {Lee},
  \citenamefont {Moore}, \citenamefont {Li}, \citenamefont {Yi}, \citenamefont
  {Hashimoto}, \citenamefont {Lu}, \citenamefont {Devereaux}, \citenamefont
  {Lee},\ and\ \citenamefont {Shen}}]{zhang2015superconducting}%
  \BibitemOpen
  \bibfield  {author} {\bibinfo {author} {\bibfnamefont {Y.}~\bibnamefont
  {Zhang}}, \bibinfo {author} {\bibfnamefont {J.~J.}\ \bibnamefont {Lee}},
  \bibinfo {author} {\bibfnamefont {R.~G.}\ \bibnamefont {Moore}}, \bibinfo
  {author} {\bibfnamefont {W.}~\bibnamefont {Li}}, \bibinfo {author}
  {\bibfnamefont {M.}~\bibnamefont {Yi}}, \bibinfo {author} {\bibfnamefont
  {M.}~\bibnamefont {Hashimoto}}, \bibinfo {author} {\bibfnamefont {D.~H.}\
  \bibnamefont {Lu}}, \bibinfo {author} {\bibfnamefont {T.~P.}\ \bibnamefont
  {Devereaux}}, \bibinfo {author} {\bibfnamefont {D.~H.}\ \bibnamefont {Lee}},
  \ and\ \bibinfo {author} {\bibfnamefont {Z.~X.}\ \bibnamefont {Shen}},\
  }\href@noop {} {\bibfield  {journal} {\bibinfo  {journal} {arXiv preprint
  arXiv:1512.06322}\ } (\bibinfo {year} {2015}{\natexlab{b}})}\BibitemShut
  {NoStop}%
\bibitem [{\citenamefont {Mikami}\ \emph {et~al.}(2002)\citenamefont {Mikami},
  \citenamefont {Nakamura}, \citenamefont {Kitao},\ and\ \citenamefont
  {Arakawa}}]{Mikami2002lattice}%
  \BibitemOpen
  \bibfield  {author} {\bibinfo {author} {\bibfnamefont {M.}~\bibnamefont
  {Mikami}}, \bibinfo {author} {\bibfnamefont {S.}~\bibnamefont {Nakamura}},
  \bibinfo {author} {\bibfnamefont {O.}~\bibnamefont {Kitao}}, \ and\ \bibinfo
  {author} {\bibfnamefont {H.}~\bibnamefont {Arakawa}},\ }\href {\doibase
  10.1103/PhysRevB.66.155213} {\bibfield  {journal} {\bibinfo  {journal} {Phys.
  Rev. B}\ }\textbf {\bibinfo {volume} {66}},\ \bibinfo {pages} {155213}
  (\bibinfo {year} {2002})}\BibitemShut {NoStop}%
\bibitem [{\citenamefont {Wang}\ \emph {et~al.}(2013)\citenamefont {Wang},
  \citenamefont {Sun}, \citenamefont {Tan}, \citenamefont {Feng}, \citenamefont
  {Cheng}, \citenamefont {Zhao}, \citenamefont {Zhao}, \citenamefont {Wang},
  \citenamefont {Luo}, \citenamefont {Yang},\ and\ \citenamefont
  {Hou}}]{wang2013role}%
  \BibitemOpen
  \bibfield  {author} {\bibinfo {author} {\bibfnamefont {Y.}~\bibnamefont
  {Wang}}, \bibinfo {author} {\bibfnamefont {H.}~\bibnamefont {Sun}}, \bibinfo
  {author} {\bibfnamefont {S.}~\bibnamefont {Tan}}, \bibinfo {author}
  {\bibfnamefont {H.}~\bibnamefont {Feng}}, \bibinfo {author} {\bibfnamefont
  {Z.}~\bibnamefont {Cheng}}, \bibinfo {author} {\bibfnamefont
  {J.}~\bibnamefont {Zhao}}, \bibinfo {author} {\bibfnamefont {A.}~\bibnamefont
  {Zhao}}, \bibinfo {author} {\bibfnamefont {B.}~\bibnamefont {Wang}}, \bibinfo
  {author} {\bibfnamefont {Y.}~\bibnamefont {Luo}}, \bibinfo {author}
  {\bibfnamefont {J.~L.}\ \bibnamefont {Yang}}, \ and\ \bibinfo {author}
  {\bibfnamefont {J.~G.}\ \bibnamefont {Hou}},\ }\href {\doibase
  10.1038/ncomms3214} {\bibfield  {journal} {\bibinfo  {journal} {Nat.
  Commun.}\ }\textbf {\bibinfo {volume} {4}},\ \bibinfo {pages} {2214}
  (\bibinfo {year} {2013})}\BibitemShut {NoStop}%
\bibitem [{\citenamefont {Fan}\ \emph {et~al.}(2015)\citenamefont {Fan},
  \citenamefont {Zhang}, \citenamefont {Liu}, \citenamefont {Yan},
  \citenamefont {Ren}, \citenamefont {Peng}, \citenamefont {Xu}, \citenamefont
  {Xie}, \citenamefont {Hu}, \citenamefont {Zhang},\ and\ \citenamefont
  {Feng}}]{fan2015plain}%
  \BibitemOpen
  \bibfield  {author} {\bibinfo {author} {\bibfnamefont {Q.}~\bibnamefont
  {Fan}}, \bibinfo {author} {\bibfnamefont {W.~H.}\ \bibnamefont {Zhang}},
  \bibinfo {author} {\bibfnamefont {X.}~\bibnamefont {Liu}}, \bibinfo {author}
  {\bibfnamefont {Y.~J.}\ \bibnamefont {Yan}}, \bibinfo {author} {\bibfnamefont
  {M.~Q.}\ \bibnamefont {Ren}}, \bibinfo {author} {\bibfnamefont
  {R.}~\bibnamefont {Peng}}, \bibinfo {author} {\bibfnamefont {H.~C.}\
  \bibnamefont {Xu}}, \bibinfo {author} {\bibfnamefont {B.~P.}\ \bibnamefont
  {Xie}}, \bibinfo {author} {\bibfnamefont {J.~P.}\ \bibnamefont {Hu}},
  \bibinfo {author} {\bibfnamefont {T.}~\bibnamefont {Zhang}}, \ and\ \bibinfo
  {author} {\bibfnamefont {D.~L.}\ \bibnamefont {Feng}},\ }\href {\doibase
  10.1038/nphys3450} {\bibfield  {journal} {\bibinfo  {journal} {Nat. Phys.}\
  }\textbf {\bibinfo {volume} {11}},\ \bibinfo {pages} {946} (\bibinfo {year}
  {2015})}\BibitemShut {NoStop}%
\bibitem [{\citenamefont {De~Wilde}\ \emph {et~al.}(1997)\citenamefont
  {De~Wilde}, \citenamefont {Iavarone}, \citenamefont {Welp}, \citenamefont
  {Metlushko}, \citenamefont {Koshelev}, \citenamefont {Aranson}, \citenamefont
  {Crabtree},\ and\ \citenamefont {Canfield}}]{Wilde1997scanning}%
  \BibitemOpen
  \bibfield  {author} {\bibinfo {author} {\bibfnamefont {Y.}~\bibnamefont
  {De~Wilde}}, \bibinfo {author} {\bibfnamefont {M.}~\bibnamefont {Iavarone}},
  \bibinfo {author} {\bibfnamefont {U.}~\bibnamefont {Welp}}, \bibinfo {author}
  {\bibfnamefont {V.}~\bibnamefont {Metlushko}}, \bibinfo {author}
  {\bibfnamefont {A.~E.}\ \bibnamefont {Koshelev}}, \bibinfo {author}
  {\bibfnamefont {I.}~\bibnamefont {Aranson}}, \bibinfo {author} {\bibfnamefont
  {G.~W.}\ \bibnamefont {Crabtree}}, \ and\ \bibinfo {author} {\bibfnamefont
  {P.~C.}\ \bibnamefont {Canfield}},\ }\href {\doibase
  10.1103/PhysRevLett.78.4273} {\bibfield  {journal} {\bibinfo  {journal}
  {Phys. Rev. Lett.}\ }\textbf {\bibinfo {volume} {78}},\ \bibinfo {pages}
  {4273} (\bibinfo {year} {1997})}\BibitemShut {NoStop}%
\bibitem [{\citenamefont {Nakai}\ \emph
  {et~al.}(2002{\natexlab{a}})\citenamefont {Nakai}, \citenamefont
  {Miranovi\ifmmode~\acute{c}\else \'{c}\fi{}}, \citenamefont {Ichioka},\ and\
  \citenamefont {Machida}}]{Nakai2002reentrant}%
  \BibitemOpen
  \bibfield  {author} {\bibinfo {author} {\bibfnamefont {N.}~\bibnamefont
  {Nakai}}, \bibinfo {author} {\bibfnamefont {P.}~\bibnamefont
  {Miranovi\ifmmode~\acute{c}\else \'{c}\fi{}}}, \bibinfo {author}
  {\bibfnamefont {M.}~\bibnamefont {Ichioka}}, \ and\ \bibinfo {author}
  {\bibfnamefont {K.}~\bibnamefont {Machida}},\ }\href {\doibase
  10.1103/PhysRevLett.89.237004} {\bibfield  {journal} {\bibinfo  {journal}
  {Phys. Rev. Lett.}\ }\textbf {\bibinfo {volume} {89}},\ \bibinfo {pages}
  {237004} (\bibinfo {year} {2002}{\natexlab{a}})}\BibitemShut {NoStop}%
\bibitem [{\citenamefont {Brown}\ \emph {et~al.}(2004)\citenamefont {Brown},
  \citenamefont {Charalambous}, \citenamefont {Jones}, \citenamefont {Forgan},
  \citenamefont {Kealey}, \citenamefont {Erb},\ and\ \citenamefont
  {Kohlbrecher}}]{Brown2004triangular}%
  \BibitemOpen
  \bibfield  {author} {\bibinfo {author} {\bibfnamefont {S.~P.}\ \bibnamefont
  {Brown}}, \bibinfo {author} {\bibfnamefont {D.}~\bibnamefont {Charalambous}},
  \bibinfo {author} {\bibfnamefont {E.~C.}\ \bibnamefont {Jones}}, \bibinfo
  {author} {\bibfnamefont {E.~M.}\ \bibnamefont {Forgan}}, \bibinfo {author}
  {\bibfnamefont {P.~G.}\ \bibnamefont {Kealey}}, \bibinfo {author}
  {\bibfnamefont {A.}~\bibnamefont {Erb}}, \ and\ \bibinfo {author}
  {\bibfnamefont {J.}~\bibnamefont {Kohlbrecher}},\ }\href {\doibase
  10.1103/PhysRevLett.92.067004} {\bibfield  {journal} {\bibinfo  {journal}
  {Phys. Rev. Lett.}\ }\textbf {\bibinfo {volume} {92}},\ \bibinfo {pages}
  {067004} (\bibinfo {year} {2004})}\BibitemShut {NoStop}%
\bibitem [{\citenamefont {Sakata}\ \emph {et~al.}(2000)\citenamefont {Sakata},
  \citenamefont {Oosawa}, \citenamefont {Matsuba}, \citenamefont {Nishida},
  \citenamefont {Takeya},\ and\ \citenamefont {Hirata}}]{Imaging2000Sakata}%
  \BibitemOpen
  \bibfield  {author} {\bibinfo {author} {\bibfnamefont {H.}~\bibnamefont
  {Sakata}}, \bibinfo {author} {\bibfnamefont {M.}~\bibnamefont {Oosawa}},
  \bibinfo {author} {\bibfnamefont {K.}~\bibnamefont {Matsuba}}, \bibinfo
  {author} {\bibfnamefont {N.}~\bibnamefont {Nishida}}, \bibinfo {author}
  {\bibfnamefont {H.}~\bibnamefont {Takeya}}, \ and\ \bibinfo {author}
  {\bibfnamefont {K.}~\bibnamefont {Hirata}},\ }\href {\doibase
  10.1103/PhysRevLett.84.1583} {\bibfield  {journal} {\bibinfo  {journal}
  {Phys. Rev. Lett.}\ }\textbf {\bibinfo {volume} {84}},\ \bibinfo {pages}
  {1583} (\bibinfo {year} {2000})}\BibitemShut {NoStop}%
\bibitem [{\citenamefont {Gilardi}\ \emph {et~al.}(2002)\citenamefont
  {Gilardi}, \citenamefont {Mesot}, \citenamefont {Drew}, \citenamefont
  {Divakar}, \citenamefont {Lee}, \citenamefont {Forgan}, \citenamefont
  {Zaharko}, \citenamefont {Conder}, \citenamefont {Aswal}, \citenamefont
  {Dewhurst}, \citenamefont {Cubitt}, \citenamefont {Momono},\ and\
  \citenamefont {Oda}}]{Gilardi2002direct}%
  \BibitemOpen
  \bibfield  {author} {\bibinfo {author} {\bibfnamefont {R.}~\bibnamefont
  {Gilardi}}, \bibinfo {author} {\bibfnamefont {J.}~\bibnamefont {Mesot}},
  \bibinfo {author} {\bibfnamefont {A.}~\bibnamefont {Drew}}, \bibinfo {author}
  {\bibfnamefont {U.}~\bibnamefont {Divakar}}, \bibinfo {author} {\bibfnamefont
  {S.~L.}\ \bibnamefont {Lee}}, \bibinfo {author} {\bibfnamefont {E.~M.}\
  \bibnamefont {Forgan}}, \bibinfo {author} {\bibfnamefont {O.}~\bibnamefont
  {Zaharko}}, \bibinfo {author} {\bibfnamefont {K.}~\bibnamefont {Conder}},
  \bibinfo {author} {\bibfnamefont {V.~K.}\ \bibnamefont {Aswal}}, \bibinfo
  {author} {\bibfnamefont {C.~D.}\ \bibnamefont {Dewhurst}}, \bibinfo {author}
  {\bibfnamefont {R.}~\bibnamefont {Cubitt}}, \bibinfo {author} {\bibfnamefont
  {N.}~\bibnamefont {Momono}}, \ and\ \bibinfo {author} {\bibfnamefont
  {M.}~\bibnamefont {Oda}},\ }\href {\doibase 10.1103/PhysRevLett.88.217003}
  {\bibfield  {journal} {\bibinfo  {journal} {Phys. Rev. Lett.}\ }\textbf
  {\bibinfo {volume} {88}},\ \bibinfo {pages} {217003} (\bibinfo {year}
  {2002})}\BibitemShut {NoStop}%
\bibitem [{\citenamefont {Zhou}\ \emph {et~al.}(2013)\citenamefont {Zhou},
  \citenamefont {Misra}, \citenamefont {da~Silva~Neto}, \citenamefont
  {Aynajian}, \citenamefont {Baumbach}, \citenamefont {Thompson}, \citenamefont
  {Bauer},\ and\ \citenamefont {Yazdani}}]{zhou2013visualizing}%
  \BibitemOpen
  \bibfield  {author} {\bibinfo {author} {\bibfnamefont {B.~B.}\ \bibnamefont
  {Zhou}}, \bibinfo {author} {\bibfnamefont {S.}~\bibnamefont {Misra}},
  \bibinfo {author} {\bibfnamefont {E.~H.}\ \bibnamefont {da~Silva~Neto}},
  \bibinfo {author} {\bibfnamefont {P.}~\bibnamefont {Aynajian}}, \bibinfo
  {author} {\bibfnamefont {R.~E.}\ \bibnamefont {Baumbach}}, \bibinfo {author}
  {\bibfnamefont {J.}~\bibnamefont {Thompson}}, \bibinfo {author}
  {\bibfnamefont {E.~D.}\ \bibnamefont {Bauer}}, \ and\ \bibinfo {author}
  {\bibfnamefont {A.}~\bibnamefont {Yazdani}},\ }\href {\doibase
  10.1038/nphys2672} {\bibfield  {journal} {\bibinfo  {journal} {Nature
  Physics}\ }\textbf {\bibinfo {volume} {9}},\ \bibinfo {pages} {474} (\bibinfo
  {year} {2013})}\BibitemShut {NoStop}%
\bibitem [{\citenamefont {Nakai}\ \emph
  {et~al.}(2002{\natexlab{b}})\citenamefont {Nakai}, \citenamefont
  {Miranovi\ifmmode~\acute{c}\else \'{c}\fi{}}, \citenamefont {Ichioka},\ and\
  \citenamefont {Machida}}]{reentrant2012Nakai}%
  \BibitemOpen
  \bibfield  {author} {\bibinfo {author} {\bibfnamefont {N.}~\bibnamefont
  {Nakai}}, \bibinfo {author} {\bibfnamefont {P.}~\bibnamefont
  {Miranovi\ifmmode~\acute{c}\else \'{c}\fi{}}}, \bibinfo {author}
  {\bibfnamefont {M.}~\bibnamefont {Ichioka}}, \ and\ \bibinfo {author}
  {\bibfnamefont {K.}~\bibnamefont {Machida}},\ }\href {\doibase
  10.1103/PhysRevLett.89.237004} {\bibfield  {journal} {\bibinfo  {journal}
  {Phys. Rev. Lett.}\ }\textbf {\bibinfo {volume} {89}},\ \bibinfo {pages}
  {237004} (\bibinfo {year} {2002}{\natexlab{b}})}\BibitemShut {NoStop}%
\bibitem [{\citenamefont {Hayashi}\ \emph {et~al.}(1998)\citenamefont
  {Hayashi}, \citenamefont {Isoshima}, \citenamefont {Ichioka},\ and\
  \citenamefont {Machida}}]{Hayashi1998low}%
  \BibitemOpen
  \bibfield  {author} {\bibinfo {author} {\bibfnamefont {N.}~\bibnamefont
  {Hayashi}}, \bibinfo {author} {\bibfnamefont {T.}~\bibnamefont {Isoshima}},
  \bibinfo {author} {\bibfnamefont {M.}~\bibnamefont {Ichioka}}, \ and\
  \bibinfo {author} {\bibfnamefont {K.}~\bibnamefont {Machida}},\ }\href
  {\doibase 10.1103/PhysRevLett.80.2921} {\bibfield  {journal} {\bibinfo
  {journal} {Phys. Rev. Lett.}\ }\textbf {\bibinfo {volume} {80}},\ \bibinfo
  {pages} {2921} (\bibinfo {year} {1998})}\BibitemShut {NoStop}%
\bibitem [{\citenamefont {Moser}\ \emph {et~al.}(2013)\citenamefont {Moser},
  \citenamefont {Moreschini}, \citenamefont {Ja\ifmmode \acute{c}\else
  \'{c}\fi{}imovi\ifmmode~\acute{c}\else \'{c}\fi{}}, \citenamefont
  {Bari\ifmmode \check{s}\else \v{s}\fi{}i\ifmmode~\acute{c}\else \'{c}\fi{}},
  \citenamefont {Berger}, \citenamefont {Magrez}, \citenamefont {Chang},
  \citenamefont {Kim}, \citenamefont {Bostwick}, \citenamefont {Rotenberg},
  \citenamefont {Forr\'o},\ and\ \citenamefont {Grioni}}]{Moser2013tunable}%
  \BibitemOpen
  \bibfield  {author} {\bibinfo {author} {\bibfnamefont {S.}~\bibnamefont
  {Moser}}, \bibinfo {author} {\bibfnamefont {L.}~\bibnamefont {Moreschini}},
  \bibinfo {author} {\bibfnamefont {J.}~\bibnamefont {Ja\ifmmode \acute{c}\else
  \'{c}\fi{}imovi\ifmmode~\acute{c}\else \'{c}\fi{}}}, \bibinfo {author}
  {\bibfnamefont {O.~S.}\ \bibnamefont {Bari\ifmmode \check{s}\else
  \v{s}\fi{}i\ifmmode~\acute{c}\else \'{c}\fi{}}}, \bibinfo {author}
  {\bibfnamefont {H.}~\bibnamefont {Berger}}, \bibinfo {author} {\bibfnamefont
  {A.}~\bibnamefont {Magrez}}, \bibinfo {author} {\bibfnamefont {Y.~J.}\
  \bibnamefont {Chang}}, \bibinfo {author} {\bibfnamefont {K.~S.}\ \bibnamefont
  {Kim}}, \bibinfo {author} {\bibfnamefont {A.}~\bibnamefont {Bostwick}},
  \bibinfo {author} {\bibfnamefont {E.}~\bibnamefont {Rotenberg}}, \bibinfo
  {author} {\bibfnamefont {L.}~\bibnamefont {Forr\'o}}, \ and\ \bibinfo
  {author} {\bibfnamefont {M.}~\bibnamefont {Grioni}},\ }\href {\doibase
  10.1103/PhysRevLett.110.196403} {\bibfield  {journal} {\bibinfo  {journal}
  {Phys. Rev. Lett.}\ }\textbf {\bibinfo {volume} {110}},\ \bibinfo {pages}
  {196403} (\bibinfo {year} {2013})}\BibitemShut {NoStop}%
\bibitem [{\citenamefont {Jiang}()}]{privatecommunication}%
  \BibitemOpen
  \bibfield  {author} {\bibinfo {author} {\bibfnamefont {P.}~\bibnamefont
  {Jiang}},\ }\href@noop {} {\bibinfo  {journal} {private communication}\
  }\BibitemShut {NoStop}%
\end{thebibliography}
%

\end{document}